\documentclass[11pt, a4paper]{article}
\usepackage[margin=25mm]{geometry}
\usepackage{coling2018}
\usepackage{times}
\usepackage{url}
\usepackage{latexsym}
\usepackage{graphicx}
\usepackage{scrextend}

\title{Triad-based Neural Network for Coreference Resolution}

\author{Yuanliang Meng\\
        Text Machine Lab for NLP\\
        Department of Computer Science \\
        University of Massachusetts Lowell \\
        {\tt ymeng@cs.uml.edu} \\\And 
        Anna Rumshisky \\
        Text Machine Lab for NLP\\
        Department of Computer Science \\
        University of Massachusetts Lowell \\
        {\tt arum@cs.uml.edu} 
}

\date{}

\begin{document}
\maketitle
\begin{abstract}
We propose a triad-based neural network system that generates affinity scores between entity mentions for coreference resolution. The system simultaneously accepts three mentions as input, taking mutual dependency and logical constraints of all three mentions into account, and thus makes more accurate predictions than the traditional pairwise approach. Depending on system choices, the affinity scores can be further used in clustering or mention ranking. Our experiments show that a standard hierarchical clustering using the scores produces state-of-art results with gold mentions on the English portion of CoNLL 2012 Shared Task. The model does not rely on many handcrafted features and is easy to train and use. The triads can also be easily extended to polyads of higher orders. To our knowledge, this is the first neural network system to model mutual dependency of more than two members at mention level.

\end{abstract}

\section{Introduction}
\label{intro}
%
%
\blfootnote{
    
    \hspace{-0.65cm}  
    This work is licensed under a Creative Commons 
    Attribution 4.0 International License.
    License details:
    \url{http://creativecommons.org/licenses/by/4.0/}
}

Entity coreference resolution aims to identify mentions that refer to the same entity. A mention is a piece of text, usually a noun, a pronoun, or a nominal phrase. 
Resolving coreference often requires understanding the full context, and sometimes also world knowledge not provided in the text. Generally speaking, three types of models have been used for coreference resolution: pairwise models, mention ranking models, and entity-mention models. The first two are more common in literature, and the third one is somewhat less studied. 

Pairwise models a.k.a. mention pair models build a binary classifier over pairs of mentions~\cite{Soon:2001:MLA:972597.972602,mccallum03}. If all the pairs are classified correctly, then all coreferent mentions are identified. The mention ranking models do not rely on the full pairwise classification, but rather compare each mention to its possible antecedents 
in order to determine whether the mention might refer to an existing antecedent or starts a new coreference chain~\cite{DurrettKlein2013,DBLP:conf/naacl/WisemanRS16,DBLP:journals/corr/ClarkM16a}. The entity-mention models try constructing representations of discourse entities, and associating different mentions with the entity representations~\cite{Luo:2004:MCR:1218955.1218973}.
Recently, some neural network models combine mention detection and coreference resolution. They design specific losses for each task, and train the components more or less jointly~\cite{P18-2017,D17-1018}. 

However, none of these model types consider more than two mentions together at the low level. By low level here, we mean the processing of input mention features, as opposed to processing of constructed representations. Pairwise models and mention ranking models make low-level decisions on mention pairs only. Some further processing may be applied to reconcile global scope conflicts, but this process no longer relies directly on mention features.

This paper proposes a neural network model which works on triads of mentions directly. Each time, the system takes three mentions as input, and decisions on their coreference relations are made while taking into account all mutual dependencies. Inferences drawn from three mentions, if correctly modeled, should be more reliable than those from two mentions, 
simply because entities in a text tend to have multiple mutual dependencies. Firstly, coreference relation is transitive, and transitivity can be revealed only by 3 or more participants. 
Secondly, mutual dependencies are not just at the level of transitivity, but can occur among lexical items, syntactic structures, or discourse information. Modeling dependency at these lower levels can therefore be helpful for coreference resolution.
We believe it is also a closer approximation of humans' cognitive process. When we read text, we often look in two or more places (including not only mentions, but also their context) to decide what a pronoun might refers to. Therefore it is reasonable to account for it at an early stage of system design. 

We show that the decisions made by the triad model are more accurate than those made by the dyad model. Such decisions can be further used in mention ranking, or simply followed up by clustering or graph partitioning as in the canonical mention pair models. 
The triad system can be easily extended to higher order polyads, if necessary. In this paper, we only consider triads, and dyads (pairs) are used for comparison. 
We use the English portion of CoNLL 2012 Shared Task dataset for training and evaluation. The original task has two parts: mention detection and coreference resolution. Our current focus is coreference, so we used the gold mentions provided in test data. 
Our experiments show that a standard hierarchical clustering algorithm using the triad model output achieves state-of-art performance under several evaluation 
measures, and it outperforms a pairwise (dyad) model. \footnote{Our source code is freely available here: \url{https://github.com/text-machine-lab/entity-coref}}

\section{Related Work}
Before the neural network models became popular in coreference resolution tasks, graphical models had often been used to capture dependencies. \newcite{mccallum03} described a system which draws pairwise inferences but also accounts for transitivity constraints.  Essentially, their model can be summarized as in equation
\begin{equation}
    P(\mathbf{y}|\mathbf{x})=\frac{1}{Z_x}\exp\left (  \sum_{i,j,l}\lambda_lf_l(x_i,x_j,y_{ij})+ \sum_{i,j,k,l'}\lambda_{l'}f_{l'}(y_{ij},y_{jk},y_{ik})\right )
\end{equation}\label{graphical}
\noindent
The first term describes the potential function of a mention pair $x_i$, $x_j$ as well as their label $y_{ij}$. For instance,  $y_{ij}=1$ if $x_i$ and $x_j$ belong to the same entity. The second term adds constraints on the labels to assure logical consistency. A  particular assignment of values to $x_i, x_j$ does not only affect the potential function involving these two nodes, but also other potential functions involving one of them. This makes the variables (mentions, in this case) dependent on each other. Exact algorithms to solve such problems are NP-hard, and some approximation techniques are often applied. 

Our proposed model can be viewed as constructing potential functions over three variables $x_i$, $x_j$ and $x_k$. However, we do not look to optimize the product of all the potential functions. Instead, we train a neural network model to assign labels to all edges within a triad locally.  Note that 
the label $y_{ij}$ for a given mention pair $x_i$, $x_j$ may have different optimal values when different $x_k$'s are used to construct a triad. The final assignment is determined by computing the average of $y_{ij}$'s. 
%
%
Moreover, our input features are mainly series of word embeddings and part of speech (POS) embeddings, encoding rich context information. The conventional algorithms used in graphical models cannot deal with such high dimensional features.


Graphical neural network (GNN) models have recently been used to process graphs~\cite{DBLP:journals/corr/DuvenaudMAGHAA15,DBLP:journals/corr/SantoroRBMPBL17,DBLP:journals/corr/Kipf}. For example, the graph convolutional networks (GCN)~\cite{DBLP:journals/corr/KipfW16,DBLP:journals/corr/DefferrardBV16} take graphs as input. Each node of the graph contains features, and a matrix represents their mutual relations. The features and the relation matrix are both used as input. Some filter layers, often shared by all nodes, process the features. The output of the filters and the relation matrix are further processed by other layers. The final output is new representations of nodes, which can be labels.

Our model shares some characteristics with GCNs. The triad input can be viewed as a basic graph: triangle, and each node is a mention. The features we used (word embeddings, POS embeddings, speaker identity, mention distance) are all associated with a node or a pair of nodes (edges). The three nodes share recurrent neural network layers. 
Because the output of such layers are used together, higher layers in the system have access to information from all the nodes. The output is a 3 dimensional binary vector, which can be considered a graphical representation too. However, our goal here is to find pairwise relations, and the triangle graphs are employed only to model (partial) mutual dependency among three mentions. 
In contrast, CGNs are capable of generating new complex representations for the nodes and they rely on the structure of the input graph, both of which are not applicable in our case.


\section{System}

The system consists of two major parts: the triad-based neural network model to compute mutual distances and a model to perform clustering. These two stages are not clearly divided, since defining mutual distances affects the clustering strategy. Generally speaking, any coreference system should have a component for local coreference, in a relatively short context, as well as a component for global coreference. We perform them in two stages, but it is also possible to build a system in which they are co-trained.

\subsection{Input Features}

Our input is triads of entity mentions. The triads have mutual (joint) features and individual features as input. Speaker identity and mention distance are mutual features. The files in the CoNLL 2012 dataset are largely transcripts of broadcast news and conversations, which typically involve several speakers. We use a binary feature to indicate whether two mentions are from utterances of the same speaker (1) or not (0). Mention distance indicates how far apart two mentions are in the text. It is the number of tokens between the start positions of two mentions. 

Individual features are word tokens and POS tokens for each entity mention. The word tokens include the mentions themselves, as well as their 8 preceding tokens and 8 succeeding tokens. We also design two special tokens to mark the beginning and end of each mention. Similarly, the POS tokens include the POS tags of the mentions, as well as the POS tags of 8 preceding and 8 succeeding tokens. Two other special tags are used to mark the beginning and end of the mentions for POS tokens too.

Each word token is represented by a 300-dimensional vector. We use \textsf{glove.840B.300d} word vectors\footnote{https://nlp.stanford.edu/projects/glove/} to initialize them, and they are updated in the training process. Each POS token is represented by a one-hot vector, and updated during training too. This enables the model to learn the similarities between different POS tags (such as NNPS and NNS, for example).
Table~\ref{tab:features} gives a summary of input features.

\begin{table}[ht]
    \centering
    \begin{tabular}{l|l}
    \hline
         \bf{Feature} &\bf{Description}\\
         \hline
         Word tokens &word embeddings of the mentions, and of 8 words before and after \\
         POS tokens &part-of-speech tag embeddings \\
         Speaker identity &whether two mentions are from the same speaker\\
         Mention distance &number of tokens between the mentions\\
         \hline
    \end{tabular}
    \caption{Input features}
    \label{tab:features}
\end{table}

\subsection{Triad Neural Network}

Word embeddings are fed into a bidirectional LSTM layer, which generates a representation for each mention. The three members of the triad share the same LSTM layer. Similarly, POS embeddings are fed into a shared bidirectional LSTM layer. 
\begin{equation}\label{word-lstm}
    h_i^{word} = \textup{Word-LSTM} (X_i^{word})
\end{equation}
\begin{equation}\label{pos-lstm}
h_i^{pos} = \textup{POS-LSTM} (X_i^{pos})
\end{equation}
\noindent
where $i=0,1,2$ is the index of the three mentions, $X_i^{word}$ is the sequence of word embeddings used to represent mention $i$, and $X_i^{pos}$ is the corresponding sequence of POS embeddings. Word-LSTM and POS-LSTM are both bidirectional, and shared by all input mentions. 
We further implement a ``mutual attention" mechanism, inspired by the alignment technique in machine translation model from \newcite{DBLP:journals/corr/BahdanauCB14}. An attention matrix $A_{ij}$ represents the attention weight from mention $j$ to mention $i$ at each time step. It is computed as:
\begin{equation}\label{equation:attention}
    A_{ij} = \textup{softmax}\left [ h_i\times h_j^T \right ]
\end{equation}
where $h_i$ and $h_j$ are the outputs of LSTM layers for mention $i$ and $j$. Word tokens and pos tag tokens are treated the same way so we use a unified symbol here for simplicity. $A_{ij}$ is an $T_i\times T_j$ matrix. $T_i$ denotes the length of the sequence for mention $i$, and similarly $T_i$ for mention $j$. In equation~\ref{equation:attention}, $h_j^T$ means the transpose of $h_j$. With this attention matrix, we can then obtain the ``context" provided by mention $j$ for mention $i$:
\begin{equation}
    C_{ij} = A_{ij}\times h_j
\end{equation}
Here $C_{ij}$ contains the information from mention $j$, needed by mention $i$. The context vector $C_{ik}$ can be computed in the same way, and then we can compute $\widetilde{h_i}$, a richer representation of mention $i$:
\begin{equation}
    \widetilde{h_i} = \textup{tanh}\left [ W_c[h_i, C_{ij}, C_{ik}]+b_c \right ]
\end{equation}
Here $W_c$ and $b_c$ are parameters to be trained. Similarly, $\widetilde{h_j}$ and $\widetilde{h_k}$ are obtained in the same way.
For each pair in the triad, the representations of two entities are concatenated with their joint features: \begin{equation}
    h_{ij} = f(X_{ij}^{speaker}, X_{ij}^{distance}, \widetilde{h_i}^{word}, \widetilde{h_j}^{word}, \widetilde{h_i}^{pos}, \widetilde{h_j}^{pos})
\end{equation}
\noindent
where $X_{ij}^{speaker}$ is a binary speaker identity feature for the mentions $i$ and $j$, $X_{ij}^{distance}$ is the positive integer feature tracking the distance between them, and $f$ represents several fully connected layer(s), shared by the three pairs. 
Our implementation uses two layers, with dropout between them. 

While $h_{ij}$ represents the relation between $i$ and $j$, the other triad member needs taken into account as well. We achieve this by constructing a shared context $h_{ijk}$:
\begin{equation} 
    h_{ijk} = g(h_{ij}\bigoplus h_{jk} \bigoplus h_{ki})
\end{equation}
\noindent
where $g$ is another fully connected layer. Operator $\bigoplus$ means elementwise vector summation. Now, we can have a decoder layer $d_{ij}$ for each of the pairwise relations.
\begin{equation} \label{repsum}
    d_{ij} = f_d(h_{ij}, h_{ijk})
\end{equation}
Function $f_d$ is another fully connected layer. The three decoders work together to generate a 3D binary vector, as in equation~\ref{output}. Each element represents whether the mention pair refers to the same entity (0 or 1). 
\begin{equation}\label{output}
    \mathbf{y} = \textup{sigmoid}(W(d_{ij}, d_{jk}, d_{ki}) + b)
\end{equation}
\noindent
where $W$ and $b$ are the weights and bias to be trained. 
The output $\mathbf{y}$ is a $1\times 3$ vector. As we can see, the three decoders do not make decisions independently, but rather, ``consult" with each other, as in equation~\ref{output}.  Each decoder also uses the shared context $h_{ijk}$ at a lower level, as seen in equation~\ref{repsum}.

\subsection{Triads Generation}
For $n$ mentions in the text, there can be ${N}\choose{3}$ triads. In most cases, we have dozens of mentions in an article, which is not an issue. However, some long articles have hundreds of mentions, so generating all triads is unpractical and unnecessary. For instance, for 100 mentions, the total number of triads would be 161,700! 

During the training process, we use only the mentions within the stretch of 15 or less. In other words, we consider the pairs with 14 or fewer mentions between them. For testing, we consider the mentions with stretches up to 40. However, this does not mean the long-distance coreference can never be detected. Often, coreferent mentions in-between the distant ones may serve as bridges, and our clustering algorithm is able to put them together. That being said, it is also true that long-distance mention pairs are less likely to corefer than those in closer proximity. Training with triads that include very distant pairs could also have the harmful effect of introducing too many negative samples.

\subsection{Dyad Baseline System}
To demonstrate that triads have advantages over a strictly pairwise approach, we also build a neural network model which takes mention pairs as input, and make binary decisions on the pairs only. The input features are the same as in the triad model, and the architecture can be considered a reduced triad system.  Now there is no context information shared by three entities. The pair representation is directly connected to the output layer. 

\section{Entity Clustering}

After the likelihood of pairwise coreference between all mentions has been determined,
we use a clustering algorithm to group them.  At the end of this process, each entity is represented by a mention cluster.

For every triad $a$, $b$ and $c$, the system will produce three real values between [0,1] to represent the ``probability" of a coreference link.  We will refer to them as affinity scores. The higher the score, the more likely a pair of mentions refers to the same entity. The affinity score over a pair is computed as the average of their scores in all triads, as shown in equation~\ref{average}.
\begin{equation}\label{average}
    \Phi(a,b)=\frac{1}{N}\sum_{c\in \mathit{W(a,b)}} \Phi(a,b;c)
\end{equation}
Here, $N$ is the total number of triads containing $(a,b)$, $\Phi(a,b)$ is the affinity score of $a$ and $b$, and $\Phi(a,b;c)$ is the affinity score of $a$ and $b$ when another mention $c$ is in the triad. $\mathit{W(a,b)}$ represents the set of mentions within the distance window of $a$ or $b$.  We have experimented with other methods besides averaging, including taking the maximum, or the average of several top candidates. We found that the average produces better results.  

The mutual distance between $a$ and $b$ is defined as the reciprocal of the affinity score, except we set the maximum value to be 10.
Since the maximum value of $\Phi(a,b)$ is 1, the minimum value of $d(a,b)$  is also 1 according to equation~\ref{distance}.  In principle, we would like distance metrics to have 0 as the minimum, which can be achieved by subtracting 1. However, for the purpose of clustering, it is not necessary.
\begin{equation}\label{distance}
    d(a, b) = min\{\frac{1}{\Phi(a,b)}, 10\}
\end{equation}
Recall that our system does not consider mention pairs too far apart in the text. For evaluation, the maximum distance for consideration is 40 (i.e. they may have up to 39 other mentions in between). We set the mutual distance between out-of-window mentions as 3.7, slightly higher than the cut-off threshold (to be explained soon). As mentioned before, this does not mean they can never be clustered together. The result depends on the choice of linkage, and whether there is any coreferent entities in-between.

We use the hierarchical clustering function provided by SciPy library
to build the sets of coreferent entities. Other than the customized distance metric, we used the default settings, opting for the $distance$, rather than $inconsistent$ cutoff criterion.
The choice of linkage has a major impact on the results. We found the $average$ linkage produces the best results. It is defined in equation~\ref{average-linkage}, where $u$ and $v$ are two clusters, $d(u,v)$ is their distance, $d(u[i], v[j])$ is the distance between cluster members $i$ and $j$, and $|\cdot|$ represents cardinalities of clusters. 
\begin{equation}\label{average-linkage}
    d(u,v)=\sum_{ij}\frac{d(u[i], v[j])}{|u|*|v|}
\end{equation}
When the clusters are built hierarchically, those with distances lower than a threshold are joined. We used $t=3.5$.

\section{Postprocessing}
The main purpose of this study is to show a triad-based model works better than a pairwise, dyad-based model. In each case, we want to keep the postprocessings minimal, and always use the same postprocessing method for both models.

First of all, we assign a minimum distance between the same proper names. In the same file, mentions of the same proper name are very likely to refer to the same entity. This may have been captured by the neural network model too, but for mentions that are very far away in the text, such a process can be useful. Secondly, we replace the pronouns ``I" and ``you" with the speaker identity, if it is available. CoNLL dataset marks the speaker for each token, if a speaker is available. Finally, we discourage clusters with pronouns only.

In a regular text, entities with pronouns only are very unlikely to occur. The neural network model and the clustering method, however, do not have any specific technique to prevent it, although the neural network should have learned to find antecedents for pronouns. After clustering is applied, we check if there is any pronoun-only clusters. If so, we pick the first mention (pronoun) in this cluster as the target pronoun, and try to find an antecedent for it. The candidates for antecedents are three non-pronoun mentions before this mention, and three non-pronoun mentions after this mention, respectively. Among all the candidates, the one with the highest affinity score with the target pronoun will be considered a possible antecedent. Then the affinity score between this possible antecedent and the target is set to 1.0, the maximum, and clustering is performed again. This method does not guarantee all pronoun-only clusters to be eliminated, because we do not want to force it. However, our analysis reveals that it does reduce pronoun-only clusters.

\section{Experiments}

For all the experiments, hyperparameters were tuned with the development set only. We use Adam optimizer with binary cross-entropy loss. The learning rate is initially set as $10^{-3}$, then $5\times 10^{-4}$ after 100 sub-epochs, and $10^{-4}$ after 100 sub-epochs. We use the term ``sub-epoch" to refer to training on 50 files, rather than the whole training set. The training set is relatively big, so we implemented a data generator with multiple subprocesses with a shared output queue. There are 1940 training files in total, so roughly all training files can be consumed in 40 sub-epochs, although smaller files may be used more frequently due to the nature of multiprocessing. The training completed in 300 sub-epochs. We use input dropout ratio 0.5 for word embeddings and POS embeddings. The last layer of each pair representations has dropout ratio 0.3.

For the baseline dyad model, the settings are similar, and the hyperparameters are as close as possible too.

\subsection{Results of Triad Model}

\begin{table}[ht]
\begin{center}
\resizebox{\textwidth}{!}{
\begin{tabular}{l| c |c c c| c c c| c c c| c}
\hline 
    & Mention &\multicolumn{3}{c|}{\bf{MUC}} &\multicolumn{3}{c|}{\bf{B$^3$}} &\multicolumn{3}{c|}{\bf{CEAF$_{\phi4}$}} & \\
    & Rec.  &Rec. &Prec. &F1 &Rec. &Prec. &F1 &Rec. &Prec. &F1 &Avg. F1 \\
\hline
chang &\bf{100.00} &\bf{83.16} &88.48 &\bf{85.74} &\bf{75.36} &79.69 &\bf{77.46} &75.38 &62.71 &68.46 &77.22\\
chen &80.82 &72.29 &89.40 &79.94 &64.60 &85.92 &73.75 &\bf{76.25} &46.40 &57.69 &70.46\\
yuan &80.03 &72.22 &89.16 &79.80 &64.75 &84.68 &73.39 &74.49 &45.46 &56.46 &69.88\\
fernandes &\bf{100.00} &70.69 &91.21 &79.65 &65.46 &\bf{85.61} &74.19 &74.71 &42.55 &54.22 &69.35\\
stamborg &78.17  &71.22 &88.12 &78.77 &64.75 &83.16 &72.81 &71.94 &43.74 &54.41 &68.66\\
\hline
Dyad model &92.73 &78.94 &87.72 &83.10 &64.42 &81.67 &72.03 &71.57 &70.20 &71.88 &75.67\\
Triad model &93.12 &81.68 &\bf{89.78} &85.54 &67.48 &83.09 &74.47 &73.91 &\bf{73.65} &\bf{73.78} &\bf{77.93}\\
\hline
\end{tabular}
}
\end{center}
\caption{\label{tab:systems} Results of coreference resolution systems on the ConLL 2012 English test data with gold mentions. Our models are the last two rows. Others' results are from \newcite{W12-4501}. After clustering, we do not force singletons to be linked, so the mention recall is not 100.}
\end{table}

Table~\ref{tab:systems} shows the results of our triad system, compared to results of other participants of the shared task. All results are on the CoNLL 2012 English test data with gold mentions. 
%

Our system performs by far the best with the CEAF$_{\phi4}$ evaluation metric, and is also near the best with the MUC metric, measured with F1 score. As a result, the averaged F1 outperforms all the participants. The top participant (chang) has a perfect mention recall, so most likely there is some mechanism to force all mentions to be linked. However, we allow singletons to exist after clustering, and they are removed before evaluation.

MUC~\cite{Vilain:1995:MCS:1072399.1072405} is a link-based metric. Mentions in the same entity/cluster are considered ``linked". MUC penalizes the missing links and incorrect links, each with the same weight. 

B$^3$~\cite{Bagga98algorithmsfor} is a mention-based metric. The evaluation score depends on the fraction of the correct mentions included in the response entities (i.e. entities created by the system). If a system does not make any decision and leaves every mention as singletons (i.e. no coreference at all), it will get a perfect precision score. \newcite{Luo:2005:CRP:1220575.1220579} indicates that the B$^3$ precision score prefers no decision. On the other hand, the recall prefers over-merging entities. We have a high B$^3$ precision and a relatively low recall, which reflects too many clusters are generated. Other systems listed in the table seem to have the same symptom. For the whole test set, there are 4532 true entities. After removing singletons for evaluation, our system generates 4548 entities. Before removing singletons, however, our system generates 5906 entities in total. On average, the clusters generated by our system have a smaller size than true entities.

CEAF$_{\phi4}$~\cite{Luo:2005:CRP:1220575.1220579} assumes each key entity should only be mapped to {\bf one} response entity, and vice versa. It aligns the key entities (clusters) with the response entities in the best way, and compute scores from that alignment. Our CEAF$_{\phi4}$ recall is similar to other participants', but the precision score is much higher-- even the top participant is over 10 points behind. It is probably because our method creates fewer entities than other systems do, after removing singletons. For all the systems, the CEAF$_{\phi4}$ scores are lower than MUC and B$^3$ scores, although the gap from our system is quite small. 


\begin{figure}[ht]
    \centering
    \includegraphics[width=1.0\textwidth]{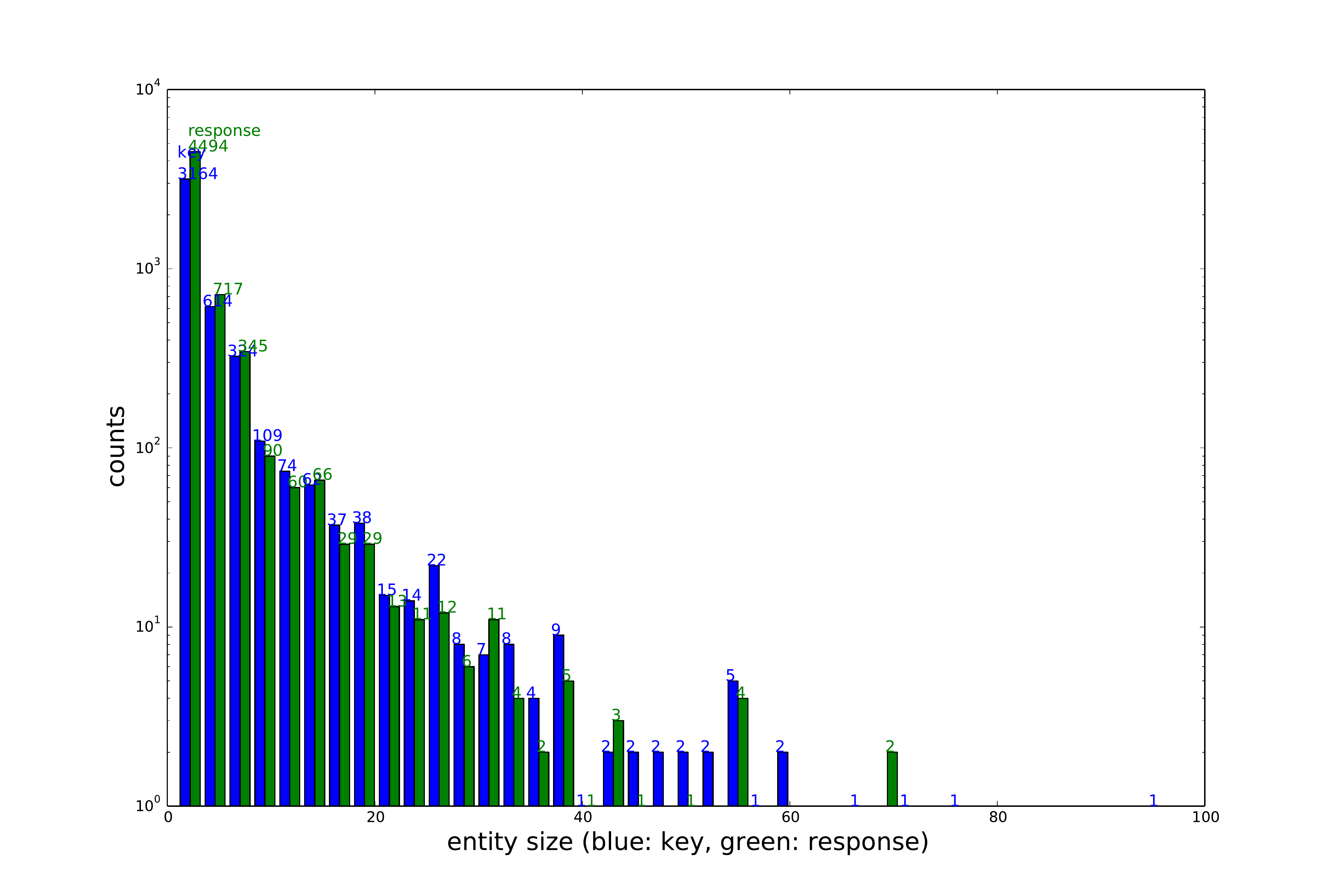}
    \caption{Entity sizes (number of mentions). Horizontal axis is the size of entities. Blue/left bars are the true counts from test set. Green/right bars are the counts from system response including singletons. Note the vertical axis is drawn in log scale. Compared to the truth, our system produces more entities with smaller sizes and misses some very large entities.}
    \label{fig:entity-sizes}
\end{figure}

In order to understand the distribution of entity/cluster sizes, we collected all the entities from test data and from our system response. Figure~\ref{fig:entity-sizes} shows the distribution of entities with respect to their size, including all the singletons our system generates. As we can see, our system generates many small entities, as shown by the left-most green bar. Note the y-axis there is in logarithmic scale, so the difference is not proportionally visualized. 

Different evaluation metrics help to diagnose different problems, and every system, as well as every metric, would have its strength and weakness. In practice, what is needed depends on the purpose.  CoNLL 2012 Shared Task uses the average of the three metrics to rank systems. 

As mentioned before, we did not perform mention detection but used the gold mentions provided in test data. However, the Shared Task required participants to perform mention detection too, and the official score is based on their detection. The top participant (fernandes) has the official average F1 score 63.7, but his average F1 on gold mentions is 69.35, not the best one. Recently, the best official score is 69.2, achieved by a combining-task neural network with 5-model ensemble~\cite{P18-2017}.

\subsection{Results of Dyads Model}
The results of the dyad system are shown in Table~\ref{tab:dyad}. The hyperparameters are chosen to be as close as possible. As we can see, the triad system has a clear advantage over the dyad system. Postprocessing also boosts the scores in either case.

\begin{table}[h!]
\begin{center}
\resizebox{\textwidth}{!}{
\begin{tabular}{l|c c c| c c c| c c c| c}
\hline 
    &\multicolumn{3}{c|}{\bf{MUC}} &\multicolumn{3}{c|}{\bf{B$^3$}} &\multicolumn{3}{c|}{\bf{CEAF$_{\phi4}$}} & \\
    &Prec. &Rec. &F1 &Prec. &Rec. &F1 &Prec. &Rec. &F1 &Avg. F1 \\
\hline

\hline
Triad	&\bf{90.38}	&80.19	&84.98	&\bf{83.86}	&64.53	&72.94	&72.01	&73.36	&72.68	&76.87\\
Dyad	&88.60	&77.82	&82.86	&82.62	&61.86	&70.75	&69.19	&71.17	&70.17	&74.59\\
Triad + post	&88.78	&\bf{81.68}	&\bf{85.54}	&83.09	&\bf{67.48}	&\bf{74.47}	&\bf{73.65}	&\bf{73.91}	&\bf{73.78}	&\bf{77.93}\\
Dyad + post	&87.72	&78.94	&83.10	&81.67	&64.42	&72.03	&70.20	&71.57	&70.88	&75.34\\
\hline
\end{tabular}
}
\end{center}
\vspace{-1em}
\caption{\label{tab:dyad} Results of the dyad model compared to the triad model. Results with postprocessing are represented with ``+ post".}
\end{table}

\begin{figure}[ht]
    \centering
    \includegraphics[width=1.0\textwidth]{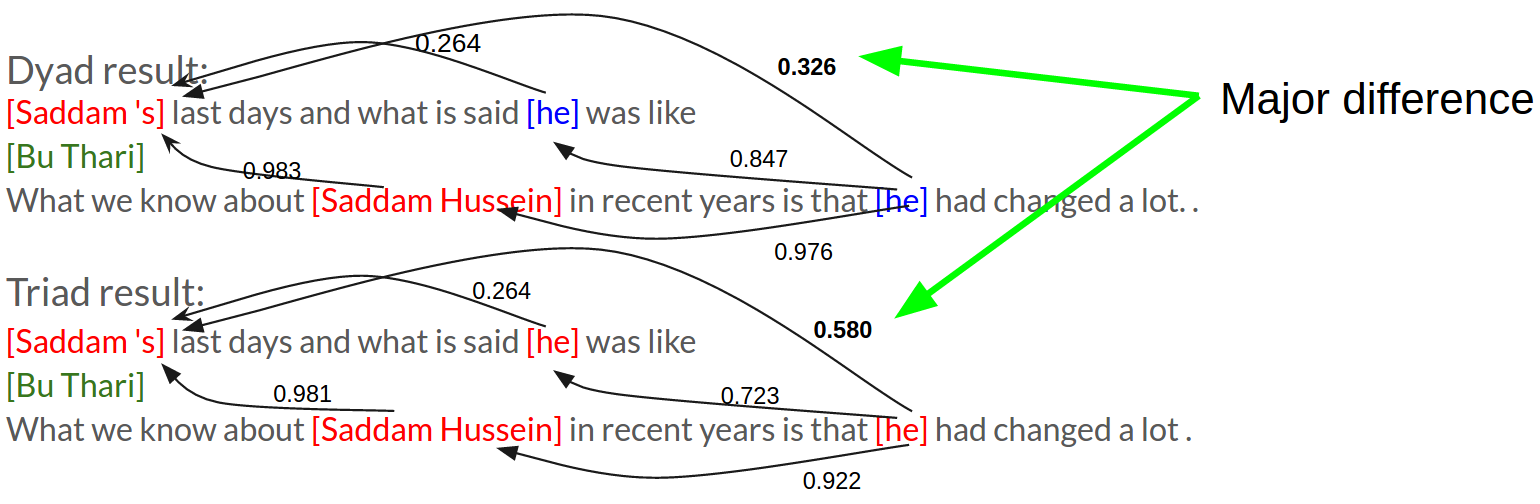}
    \caption{The affinity scores from dyad and triad models. }
    \label{fig:comparison}
\end{figure}

The following example in Figure~\ref{fig:comparison} illustrates the different results from the two systems. {\it Saddam's} and the second {\it he} should corefer. The dyad model assigns a relatively low affinity score 0.326, but the triad model assigns a much higher score 0.580. As a result, the dyad model fails to build the coreference relationship after clustering while the triad model succeeds.
With a closer look, we find that the triad with the other mention {\it Saddam Hussein} is the most helpful. With that mention as the third member in a triad, the affinity score between {\it Saddam's} and the second {\it he} reaches 0.830. Triads with other mentions, i.e. the first {\it he} or the author name {\it Bu Thari} yield near-neutral scores for this pair, in the 0.4$\sim$0.5 range.

Triad model can also support additional restrictions.
For example, we can require at least one pair in a triad to have a short distance in the text.
The point of allowing longer distances between mentions is to identify coreferent mentions that are far apart in text. 
However, it is typically fairly rare to have mentions that are far away refer to the same entity.
We do not have to allow all sides of a triangle to be big, and imposing this restriction may improve the overall quality of the response entities.

Note that this system can be easily extended from triads to tetrads (union of four mentions) and higher polyads.
Sometimes we may want to look at two more other places to determine whether a coreference relation is present. Ideally, the larger the polyad, the better we can capture mutual dependencies. However, since the number of polyads grows fast with the polyad order, the computation may quickly become intractable for larger texts.


\section{Conclusion}
We developed a triad-based neural network model that assigns affinity scores to mention pairs. A standard clustering algorithm using the resulting scores produces state-of-art performance on gold mentions. Particularly, our systems achieves much better CEAF$_{\phi4}$ F1 score.
A dyad-based baseline model has lower performance, suggesting that using triads plays an important role.  
Note that approaches other than clustering, such as the mention ranking models, can easily be used with our output as well, and we expect some of them would work better than the simple agglomerative clustering.


Mutual dependencies among multiple mentions are important in coreference resolution tasks, but it is often ignored. Our triad-based model addresses this gap.  
This model can be additionally constrained to improve performance, and if necessary, easily extended from triads to polyads with higher order.

\section*{Acknowledgments}
This project is funded in part by an NSF CAREER award to Anna Rumshisky (IIS-1652742).


\bibliographystyle{acl}
\bibliography{coling2018}

\end{document}